\documentclass[epsfig,12pt]{article}
\usepackage{epsfig}
\usepackage{graphicx}

\newcommand{\beq}{\begin{equation}}   
\newcommand{\eeq}{\end{equation}}

\newcommand{\gsim}{\lower.7ex\hbox{$
\;\stackrel{\textstyle>}{\sim}\;$}}
\newcommand{\lsim}{\lower.7ex\hbox{$
\;\stackrel{\textstyle<}{\sim}\;$}}

\begin{document}
\begin{flushright}
FTPI-MINN-13/25, UMN-TH-3215/13\\
July 22, 2013
\end{flushright}

\vspace{0.3cm}

\begin{center}
{\Large Remarks on Adjoint QCD with $k$ Flavors, $k\geq 2$}.

\vspace{3mm}

{\large M. Shifman}

\vspace{0.5cm}
{\em William I. Fine Theoretical Physics Institute, University of Minnesota,
Minneapolis, MN 55455, USA}

\vspace{0.5cm}

{\bf Abstract}

\end{center}

\vspace{0.3cm}

{\small  I summarize what we know of adjoint QCD. Some observations (albeit very simple) are new. }

\section*{Introduction}

Currently we observe a certain revival of interest to QCD with fermions in the adjoint representation of the
SU$(N)$ gauge group, the so called  ``adjoint quarks." This is due to a provocative claim \cite{un}
that this theory (to be referred to as adjoint QCD, or AQCD) with {\em more than one flavor} ($k\geq 2$), being 
non-supersymmetric at the Lagrangian level,
develops a supersymmetric spectrum of color-singlet hadrons at $N=\infty$, possibly, with the exception of a few low-lying states. Supersymmetric spectrum is defined as follows \cite{un}: each color-singlet hadron with integer spin (boson) is accompanied
by a hadron with half-integer spin (fermion) of exactly the same mass $M$ (in the limit $N=\infty$). The overall number of bosonic color-singlet degrees of freedom at each level $M$ matches that of fermionic degrees of freedom, with a possible exception of several states whose number does not grow with $M$.

I revisit adjoint QCD
with the goal to summarize  what we know about AQCD with $k\geq 2$, to see whether or not such scenario is possible. 
No indications on infrared (IR) large-$N$ ``accidental" supersymmetry are detected. Moreover, at $k=5$ 
adjoint QCD is in the conformal regime in the IR, with small anomalous dimensions,
and the number of the fermion degrees of freedom certainly does not match the number of the
boson degrees of freedom. Most probably, the  conformality extends to $k=4$. At $k=2$
one observes a certain symmetry breaking pattern for a continuous chiral symmetry, with two massless ``pions" emerging as a result of this breaking. At the same time, massless pions certainly cannot appear at weak coupling. This implies a phase transition in passage from weak to strong coupling. 

\section*{Coupling constant}

The Lagrangian of AQCD has the form
\beq
{\mathcal L} = -\frac{1}{4g^2}\, G_{\mu\nu}^a G^{\mu\nu,a}+\sum_k \left(\bar\lambda^a_{\dot\alpha}\right)_k
i{\mathcal D}^{\dot\alpha\alpha} \left(\lambda^a_{\alpha}\right)^k
\label{lagr}
\eeq
where $k$ can be $1,2,3,4$ or 5. If $k>5$, asymptotic freedom is lost, the theory becomes IR free and uninteresting.
The indices $\alpha, \dot\alpha$ in (\ref{lagr}) are Lorentz-spinorial, $a$ is the index of the adjoint 
representation of SU$(N)$, and $g^2$ is the gauge coupling,
$$\alpha \equiv \frac{g^2}{4\pi}\,.$$
If I write, say, $\lambda$ without the adjoint index,
this will mean $$\lambda\equiv\lambda^a T^a$$
where $T^a$ stand for the generators of SU$(N)$ in the fundamental representation.

If $k=1$ the Lagrangian (\ref{lagr}) represents ${\mathcal N}=1$ supersymmetric Yang-Mills (SYM) theory.
Needless to say that supersymmetry is exact not only at the Lagrangian level, but in the hadronic spectrum and scattering amplitudes too. We will discuss $k=2$ or larger.

The two-loop $\beta$ function of AQCD with arbitrary $k$ can be extracted from \cite{tlb},
\beq
\beta = \mu\,\frac{\partial}{\partial\mu}\, \alpha(\mu) \equiv -\beta_0\,\frac{\alpha^2}{2\pi}+\beta_1\, 
\frac{\alpha^3}{2(2\pi )^2}\,,
\label{dva}
\eeq
where
\beq
\beta_0= \left(\frac{11}{3} -\frac{2}{3}\, k\right)N\,,\qquad \beta_1=\left(-\frac{34}{3} + \frac{16}{3}\, k\right)N^2\,.
\eeq

At $k=5$ AQCD, still being asymptotically free, develops an IR fixed point in the perturbative domain, at
\beq
\frac{N\alpha_*}{4\pi} = \frac{1}{46}\,. 
\label{tri}
\eeq
Indeed, at $k=5$ the value of $\beta_0$ is abnormally small, while the value of $\beta_1$ is {\em positive}
and is of the normal order of magnitude. As a result, the $\beta$ function (\ref{dva}) has a reliable zero at a small value of
$\alpha$, see (\ref{tri}). The latter is smaller than the value of the QCD coupling constant $\alpha_s$  at the $Z$ peak, where perturbation theory is applicable beyond any doubt. Thus, here we encounter the regime
first described by \cite{bm,bz} which goes under the name of the Banks-Zaks phenomenon. The above IR fixed point implies
conformally symmetric theory in the infrared, with small anomalous dimensions. Neither confinement nor spontaneous breaking of the chiral symmetry ($\chi$SB, see below) are implemented.

The same probably applies to four flavors, $k=4$, too. Indeed, the would-be position of the zero of the $\beta$ function is
given below, in the upper line,
\beq
\frac{N\alpha_*}{4\pi} = \left\{\begin{array}{l}
 \frac{1}{10}\,,\qquad k=4,\\[2mm]
  \frac{5}{14}\,,\qquad k=3,
\end{array}
\right.
\label{5five}
\eeq
while nothing can be said about the existence (or nonexistence) of the IR fixed point at $k=3$. At $k=2$ the coefficient 
$\beta_2$ becomes negative.\footnote{Beyond two loops the
coefficients of the $\beta$ function are scheme-dependent. In the first and second loop only planar graphs
contribute. This is not the case in higher orders,  and e.g. at the four-loop order  the right-hand side would contain $1/N^2$
corrections. Calculations including the third and fourth loops in a reasonable scheme   performed
in \cite {hl} indicate that the actual value of $\frac{N\alpha_*}{4\pi} $ for $k=4$ is in fact somewhat lower than that
indicated in (\ref{5five}) enhancing the probability that $k=4$ belongs to the conformal window. The same conclusion is supported by lattice data \cite{ld}. 

It would be very interesting to reliably determine whether the left edge of the conformal window lies at $k=3$ or $k=4$. Note that in the model at hand, if we limit ourselves to the first and second loops, 
both edges of the conformal window depend only on the number of flavors, rather than on the ratio $N_f/N_c$
as in Seiberg's supersymmetric conformal window \cite{sei}. It is worth adding that in three-color QCD with $N_f=15$
widely believed to be conformal, $\frac{N\alpha_*}{4\pi}= \frac{3}{88}\sim \frac{1}{30}$. In the Intriligator-Seiberg-Shenker model (supersymmetric SU(2) Yang-Mills with the chiral quark field in the 3/2 representation of 
SU(2)), which was
 argued to be conformal \cite{unp,we}, $\frac{N\alpha_*}{4\pi}= \frac{2}{75}\sim \frac{1}{40}$
\cite{we}.
} 

AQCD with $k=1,2$ is believed to be confining. Thus, in our discussion we focus on these two cases. 

\section*{Chiral properties}

Let us start from SYM theory which is a particular case of AQCD with $k=1$. The only fermion current in this theory is
\beq
R^\mu =\bar\lambda_{\dot\alpha}^a\left(\bar\sigma^\mu\right)^{\dot\alpha\alpha}
\lambda_{\alpha}^a
\,.
\label{six}
\eeq
Classically it is conserved; however, at the quantum level it is internally anomalous,
$$
\partial_\mu R^\mu = \frac{N}{16\pi^2}\, G_{\mu\nu}^a \tilde G^{\mu\nu,a}\,.
$$
Moreover, $\partial_\mu R^\mu\neq 0$ even in the limit $N\to \infty$, in contradistinction with the singlet fermion current in QCD.

Thus, in $k=1$ AQCD there is no conserved fermion charge. However, the conserved operator
$(-1)^F$, distinguishing fermions from bosons can be introduced. Indeed, the classical conservation of the current (\ref{six}) leaves a remnant in the form of the discrete 
$Z_{2N}$
symmetry which is dynamically broken down \cite{w1,w2} to $Z_2$ by the gluino condensate 
$\left\langle{\rm Tr}\lambda^2\right\rangle\neq 0$ (for more details see \cite{uche}).
The existence of this operator is needed e.g. for the Witten index determination \cite{w1}. 
Considering a given color-singlet (hadronic) state, say, bosonic, one cannot say, however,
whether it contains zero, two, four, six and so on ``fermion quanta." In this sense, unlike QCD in  which the 
flavor charge is well defined, the notion of a ``constituent" quark in SYM theory is meaningless. The operators ${\rm Tr}\lambda^2$
and ${\rm Tr}\,G^2$ have the same value of $(-1)^F$, while the operators ${\rm Tr}\,\lambda^2$
and ${\rm Tr}\,\lambda G $ have the values $+1$ and $-1$, respectively, and produce degenerate spin-0 and spin-1/2 states.

In $k=2$ AQCD the chiral symmetry of the Lagrangian is SU(2). Indeed, the Lagrangian stays invariant under arbitrary rotations
\beq
\left(\begin{array}{l}
\lambda^1\\[1mm]
\lambda^2
\end{array}
\right)\to U\, \left(\begin{array}{l}
\lambda^1\\[1mm]
\lambda^2
\end{array}
\right)\,, \qquad U\in {\rm  SU}(2)\,,
\label{sev}
\eeq
where the superscripts 1 and 2 denote the value of the flavor index $k$. In addition, at the classical level there exists the 
U(1) symmetry generated by the current (\ref{six}) with summation over two flavors. This symmetry is anomalous; there is no need to consider it here.\footnote{As in supersymmetric Yang-Mills theory (or, which is the same, $k=1$ AQCD), the anomalous current $R^\mu =\left( \bar\lambda_{1}^a\, \bar\sigma^\mu 
\lambda_{\alpha}^{a,1} + \bar\lambda_{2}^a\, \bar\sigma^\mu 
\lambda_{\alpha}^{a,2}\right)$  still can be used to define an operator $(-1)^f$ \cite{bol}.
The eigenvalue of $(-1)^f$ is $1$ for any operator with the odd number of the $\lambda$ and $\bar\lambda$ fields
and $-1$ for any operator with the even number of the $\lambda$ and $\bar\lambda$ fields. Moreover, $k=2$ AQCD
 supports topologically stable solitons with mass scaling as $N^2$ \cite{bol}, see below. Topological stability is due to the existence of a nontrivial Hopf invariant in the Skyrme-Faddeev model. All ``normal" hadrons, with mass $O(N^0)$, are characterized by $(-1)^f (-1)^F =1$, while for the Skyrmion states with mass $O(N^2)$  the value
 of $(-1)^f (-1)^F = -1$.
} 

Needless to say, the chiral symmetry (\ref{sev}) is spontaneously broken. The pattern of this breaking can be
obtained\footnote{For $k=5$ (and, probably, $k=4$)
the 't Hooft matching should be trivial since there is no confinement, see above.} from the 't Hooft matching \cite{DZ2}.  Assuming confinement and the large-$N$ limit, as it had been done in
\cite{colewit}, one can conclude that
\beq
{\rm  SU}(2)\to {\rm  U}(1)\,.
\label{ei}
\eeq
The corresponding analysis in AQCD with arbitrary $k$ was carried out in \cite{ks}. The order parameter triggering the above $\chi$SB can be chosen as follows:
\beq
\left\langle{\rm Tr}\lambda^1\lambda^2\right\rangle + (1\leftrightarrow 2) \neq 0
\label{ni}
\eeq
with the same convention on superscripts as in Eq. (\ref{sev}). Then the conserved unbroken U(1) current has the form
\beq
j_{{\rm U}(1)}^\mu=  2\left( {\rm Tr} \, \bar\lambda_1 \bar\sigma^\mu
\lambda^1 - {\rm Tr} \, \bar\lambda_2 \bar\sigma^\mu
\lambda^2
\right)\,.
\label{ten}
\eeq
It generates rotations of two Weyl spinors $\lambda^{1,2}$ in the opposite direction.
We could have rewritten $k=2$ AQCD as a gauge theory of a single adjoint Dirac spinor. Then the
current (\ref{ten}) will obvious become the vector fermion current. Thus, in this theory the fermion charge $F$ is
perfectly defined,
\beq
F=\int d^3 x\, j_{{\rm U}(1)}^0\,,
\label{efff}
\eeq
 unlike the SYM theory.
 
 The $\chi$SB pattern (\ref{ei}) gives rise to two massless ``pions" coupled to two broken currents; the U(1) charges of these pions are $+2$ and $-2$, respectively. 
 
 Note that $k=2$ AQCD considered on the small-$L$ cylinder as in \cite{un} (i.e. at weak coupling) can never produce
massless pions. Thus, as we change $L$ from $L\ll \Lambda^{-1}$ to $L\gg \Lambda^{-1}$ (strong coupling) a phase transition is inevitable.

The existence of the conserved charge $F$ in $k=2$ AQCD splits the Hilbert space of physical states (hadrons) into sectors with the given value of $F$,
$$ F=0, \pm 1, \pm 2, ...$$
This allows for a meaningful introduction of a ``constituent" adjoint quark. Producing two extra adjoint quarks moves us from the $F=0$ sector  to the $F=2$ sector, from $F=1$   to  $F=3$, and so on.  The operators ${\rm Tr}\lambda^1\lambda^1$
and ${\rm Tr}\,G^2$ become distinguishable, and so are  ${\rm Tr}\bar\lambda_2\lambda^1$
and ${\rm Tr}\,G^2$. At the same time ${\rm Tr}\lambda^2\lambda^1$
and ${\rm Tr}\,G^2$ both have vacuum quantum numbers.

Returning to the chiral properties, I should mention that the chiral symmetry breaking and the emergence of two pions have an impact not only on the low-lying states, but on high excitations too. The linear realization of the chiral symmetry is not  restored in the highly excited states, unless the Regge trajectories  intersect, which is unlikely  \cite{msav}.
If so, the Goldberger-Treiman relation should take place for (infinitely many) spin-1/2 states. In addition, the mesons and baryons forming chiral pairs must be split (non-degenerate in masses).

In the hadronic spectrum there are infinitely many sectors characterized by $F=0,\pm 1, \pm 2,...$. 
At $N=\infty$ not only the lowest-lying states in each sector (these sectors extend all the way up to the Skyrmion sector) are stable, but so are all excitations.

\section*{Multiquark states in AQCD}

In multicolor QCD (i.e. in the 't Hooft limit \cite{thooft}) exotic mesons with more than one quark-antiquark pair are not bound and split
into a number of noninteracting nonexotic mesons, each of which contains exactly one $q\bar q$
pair connected by a gluon string
\cite{thooft,w3}. The string does not break at $N=\infty$.

This is {\em not} the case in $k=2$ AQCD. It is easy to see that unbreakable color-singlet states with as many quarks
as one wants do exist. For instance, consider the string operator
\begin{eqnarray}
&& {\rm Tr}\left[
 \lambda^1(x_1)\, \exp\left(i\int_{x_2}^{x_1} dx_\mu A^\mu (x)
\right)\bar\lambda_2 (x_2) \, \exp\left(i\int_{x_2}^{x_3} dx_\mu A^\mu (x )
\right) \right.
\nonumber\\[3mm]
&&\times \left.\lambda^1(x_3)\, \exp\left(i\int_{x_3}^{x_4} dx_\mu A^\mu (x)
\right) \lambda^1(x_4)\, \exp\left(i\int_{x_4}^{x_1} dx_\mu A^\mu (x)
\right)
\right]
\label{twelve}
\end{eqnarray}
(see Fig. 1). It has $F=4$ and, at the same time,  cannot be split into a product of two color-singlet operators with $F=2$
each.
I will return to this point later in the context of {\em planar equivalence} \cite{pe1,pe2}.

\begin{figure}[h]
  \begin{center}
  \includegraphics[width=2.5in]{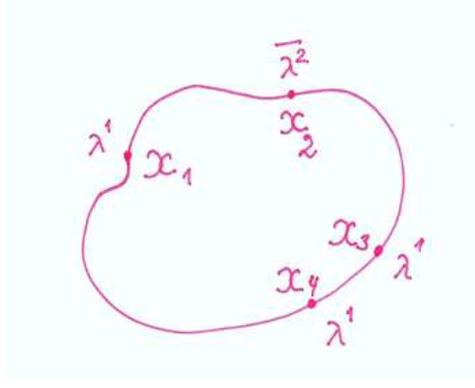}
  \caption{\small Graphic representation for the integration contour in the
  operator (\ref{twelve}).}
   \end{center}
\label{aaa}
\end{figure}

In pure Yang-Mills theory ($k=0$) all color-singlet states are represented by excitations of a closed string, which can be written as the following Wilson operator:
\beq
W_C = \frac{1}{N} {\rm Tr}\, \exp \left(i\int_C dx_\mu A^\mu (x)
\right)
\eeq
where the integration contour can be chosen, for instance, as in Fig. 1. In QCD the ``meson" 
string must be open, with (anti)quarks attached to its endpoints (Fig. 2). Glueballs are still produced
 by closed strings.

\begin{figure}[h]
  \begin{center}
  \includegraphics[width=2in]{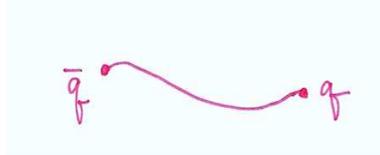}
  \caption{\small Open string corresponding to mesons in
  QCD with fundamental quarks.}
   \end{center}
\label{bbb}
\end{figure}

Since the notion of the constituent quark is well-defined in $k=2$ AQCD, it is natural to expect that
the mass of a given hadron for not too large and fixed angular momentum $L$ (i.e. $L\lsim F$) depends on $F$ as follows
\beq
M_F = \Lambda\, (a + b F)\,,
\label{14}
\eeq
where $a$ and $b$ are $F$ independent constants. At the very least, I would say
that in the large-$F$ limit $\frac{\partial M_F}{\partial F}$ = const  is practically unavoidable. The color-singlet hadrons in the model at hand
resemble nuclei in conventional QCD. 

\section*{Planar equivalence}

Because of the planar equivalence \cite{pe1,pe2} one can relate AQCD  to 
its orientifold daughter: Yang-Mills theory with Dirac fermions $\psi$, each in the two-index antisymmetric representation of
the color SU$(N)$ group,
\beq
\lambda_{\alpha, j}^i \leftrightarrow \{\chi_{\alpha}^{ij}\,,\,\,\eta_{\alpha , ij}
\}\,,\qquad \psi^{ij} = \{\chi_{\alpha}^{ij}\,,\,\,\bar\eta^{\dot\alpha,ij}
\}\,.
\eeq
For $k=2$ it is convenient to define two Dirac fermions of the daughter theory as follows:
\beq
\psi^1 =\left(\begin{array}{l}
\chi^1\\
\bar\eta_2
\end{array}
\right),\qquad \psi^2 =\left(\begin{array}{l}
\chi^2\\
\bar\eta_1
\end{array}
\right).
\eeq
\begin{figure}[h]
  \begin{center}
  \includegraphics[width=2.5in]{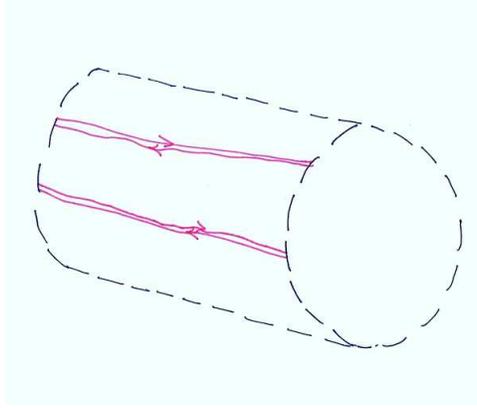}
  \caption{\small A fragment of the closed string world sheet. 
  Double line paths present the world lines of the $\lambda$ insertions. }
   \end{center}
\label{ccc}
\end{figure}
Then 
\beq
{\rm Tr}\lambda^1\lambda^2 + (1\leftrightarrow 2) + {\rm h.c.}\to \bar\psi_1\psi^1 + \bar\psi_2\psi^2\,.
\label{nip}
\eeq
The conserved U(1) current from Eq. (\ref{ten}) takes the form
\beq
j_{{\rm U}(1)}^\mu= \left(  \bar\psi_1 \gamma^\mu
\psi^1 -  \bar\psi_2 \gamma^\mu
\psi^2
\right)\,.
\label{tenp}
\eeq

The equivalence holds in the common sector. At $N=\infty$ three domains of the cylinder depicted in Fig. 3 become dynamically disconnected. To pass from $k=2$ AQCD to the daughter theory one must cut out the middle sector, flip it around the vertical axis, and glue back. The opposite arrows indicating color flow on the fermion lines become aligned. In this passage terms of the relative order $1/N$ must be ignored. 

As was mentioned in \cite{pe1,pe2}, the two-index antisymmetric Dirac fermions present a different way of the large-$N$ 
extension of {\em bona fide} QCD, sometimes referred to as the ASV continuation (different compared to 't Hooft's continuation \cite{thooft}, with fermions in the fundamental for any $N$). 
At $N=3$ both are equivalent, since at $N=3$ the two-index antisymmetric quark is  exactly the same as the fundamental antiquark.
The 't Hooft line of reasoning predicts that exotic (multiquark) mesons do not exist at $N=\infty$. The ASV procedure, with the Wilson operator defined in (\ref{twelve}) and conserved fermion number $F$, yields stable mesons  with arbitrary $F$ in the limit $N=\infty$.

Certainly, $N=\infty$ is not the same as $N=3$. However, in other aspects of phenomenology 
both alternative continuations -- that of 't Hooft and the ASV  procedure -- lead to results of comparable quality \cite{Cherman}, even for baryons. One can view this fact as an indication that multicolor QCD generally speaking does {\em not} disfavor exotic or cryptoexotic (four-quark) mesons
in {\em bona fide} QCD. They are likely to be implemented in the form of a bound diquark-antidiquark pair \cite{Selem, Wilczek,Shifman}. Needless to say, there are no traces of supersymmetry in multicolor 
QCD.

\section*{Regge trajectories}

Noncritical string theory describing a ``real" pure Yang-Mills theory in four dimensions does not exist, let alone
Yang-Mills theory with fermions, such as SYM theory or QCD with massless quarks and $\chi$SB.
Therefore, exact predictions for the Regge trajectories are unavailable. There are all reasons to hope, however, that for large excitation numbers the quasiclassical approximation for the Regge trajectories must work well. Quasiclassical calculations
reproduce the famous Chew-Frautschi formula \cite{cf}, with the linear dependence of the meson 
and baryon masses {\em squared} on the angular momentum $L$
and the excitation number $n$ (the so-called primary and daughter Regge trajectories). In fact, the linear dependencies are clearly seen in experiment even for the
lowest-lying states (see e.g. \cite{msav}) in all cases where data are available, with the exception of the Pomeron trajectory.

If Eq. (\ref{14}) is valid I do not expect linear Regge trajectories in AQCD for high-lying states,
$F,L\gg1$ (but $L\lsim F$) because of the interplay of the linear in mass dependence in (\ref{14}) and quadratic
in mass  in the Chew-Frautschi formula.

\section*{A parallel (perhaps, rather remote)}

Equation (\ref{twelve}) with $\lambda$ insertions in the closed loop, non-factorizable at $N=\infty$, resembles 
the construction worked out in \cite{SY,T} where confined monopoles were identified as kinks in the string world-sheet theory. Then the $\lambda$ insertion in the supersymmetric case ($k=1$) can be viewed as a massless ``kink" while
 at $k=2$ the kinks acquire a mass.

\section*{Conclusion}

Physics of AQCD is such that
the claim \cite{un} seems implausible, although this conclusion -- I must admit -- is   not at the level of a mathematical theorem.

\section*{Acknowledgments}

I am grateful to A. Armoni, A. Cherman, T. DeGrand, R. Shrock,  M.~\"Unsal, and A. Vainshtein  for useful communications. 

This work  is supported in part by DOE grant DE-FG02-94ER40823.

\end{document}